\pdfoutput=1 
\documentclass{JINST}
\usepackage{subfig}
\RequirePackage{lineno}
\usepackage{lineno}

\title{Developing Scintillation Light Readout Simulation for the SBND
 experiment}

\author{D. Garcia-Gamez$^a$\thanks{Corresponding author.}\\
\llap{$^a$}The University of Manchester,\\
  School of Physics and Astronomy, United Kingdom\\
E-mail: \email{diego.garciagamez@manchester.ac.uk}}

\abstract{Detection of scintillation light can play several important
  roles in LArTPCs. Increased collection efficiency could result in
  the improvement of time, energy, and position resolution. The SBND
  collaboration is developing detailed MC simulations to study the
  performance of different types of light systems in the LArSoft
  framework. Due to the vast  number of photons typically produced in
  neutrino physics events, a full optical simulation becomes extremely
  hard to run on reasonable time scales. I will describe how the SBND
  simulation tackles these problems and its current status for two of
  the light detection systems considered by SBND: (i) a traditional
  TPB-coated PMT based system and (ii) a system based on TPB-coated
  reflector foils to increase collection efficiency without increasing
  the number of photodetectors. } 


\begin{document}
\section{Introduction}\label{sec:introduction}

Free electron separation and
VUV light emission are the two features that characterize the use of
liquid argon (LAr) as active medium~\cite{doke}. Both processes are complementary and their
relative weight depends on the actual strength of the applied electric
field. The free electron yield rises with the field value while the
photon yield decreases due to a reduction of the recombination
rates. At the typical field strength value of 500$\,V/cm$ both
processes (ionization and scintillation) are relevant. The free
electrons from ionization induce detectable signals on the wires of
the Time Projection Chambers (TPC) during their drift motion towards
and across the wire planes.

Depending on the performance of the light detection system (LDS), the
detection of scintillation light can provide important
information. Increasing collection efficiency results in the
improvement of time, position and energy resolution, as well as in a
lower threshold for the energy reconstruction. For a detector in a
beam (as SBND~\cite{sbn}),  prompt light signals can provide internal trigger
formation, tagging events in phase with the neutrino beam. It can
provide an effective method for an absolute time measurement, needed
for the reconstruction of non-beam related events such as cosmic rays
or supernova neutrinos.  The light signals contain additional
(complementary) information for total energy deposition reconstruction
and for particle identification. There are additional opportunities
offered by enhanced LDSs, for example, it is possible to determine the
sign of the incoming neutrino through tagging of Michel electrons
coming from stopping muons~\cite{michel}.

\section{Components of the Light Detection System in
  SBND}\label{sec:ldsSBND}

An important part of the mission of SBND is to be an R$\&$D for future LAr neutrino experiments. Its relatively small
volume makes it an excellent test-bed for new light detection system
designs. The LDS components under evaluation for SBND include:
traditional TPB-coated PMTs, TPB-coated reflector foils, and  acrylic light
guide bars read out at the ends with SiPMs. In this proceeding I will
focus on the two first components mentioned above.
\paragraph{TPB-coated PMTs.} It has been demonstrated that a system
based on TPB-coated cryogenic PMTs works well in modest-sized
experiments (ICARUS~\cite{icarus}, MicroBooNE~\cite{sbn}), but they are
difficult to scale to bigger ones (like DUNE).\\
\textbf{TPB-coated reflector foils.} Adapted from liquid argon dark
matter detectors, the installation of these foils inside the time projection chamber
volume enhance light collection without increasing the number of
photodetectors. A system of this type is currently implemented in the
LArIAT experiment~\cite{lariat}.  

In this work we present the methods being developed to benchmark these
different components. In particular studying the effects of adding
wavelength shifter covered reflector foils 
to an array of 3'' diameter PMTs (154 units).
In the next sections we will describe the workings of these
simulations, and compare the first results,
regarding the detection efficiency, obtained for the systems under
evaluation.

\section{The optical simulation in LArSoft}\label{sec:larsoft}

SBND simulations are performed within the open source LArSoft
framework, which provides simulation reconstruction and analysis
tools for current and future LAr TPC experiments. Basically, different
detectors only require their own geometry definitions (in gdml format)
and a small number of specific detector settings. 

The simulation of the optical photons (production and propagation) in
LArSoft incorporates two methods~\cite{sim}. The \textbf{full optical simulation}
implements the production and tracking of individual scintillation
photons using Geant4. To produce a realistic detector response to the
generated light, Rayleigh scattering, reflections, wavelength shifting and
absorption are considered in the tracking of light. The
huge number of photons typically produced in a neutrino physics event
makes these simulations extremely slow and CPU consuming, taking on the
order of hours or even days per event. An alternative \textbf{fast optical
  simulation} mode has been developed to overcome this problem for
regular simulation tasks. This approach is based in the existence of a
previously full-mode-built library of stored visibility data (visibility $\equiv$
ratio between the number of produced and detected photons) to
sample an expected detector response given an isotropic emission of
light at some point in the active volume. Thus, for each energy
deposition at each step, instead of generating Geant4 trackable optical
photons, the fast scintillation mode predicts a certain number of
these photons arriving to each optical sensitive volume. With this
procedure, the simulation of an event typically takes minutes rather
than hours to finish. 

\subsection{Optical library generation}

\begin{figure}[tbp] 
\centering
\includegraphics[width=0.5\textwidth]{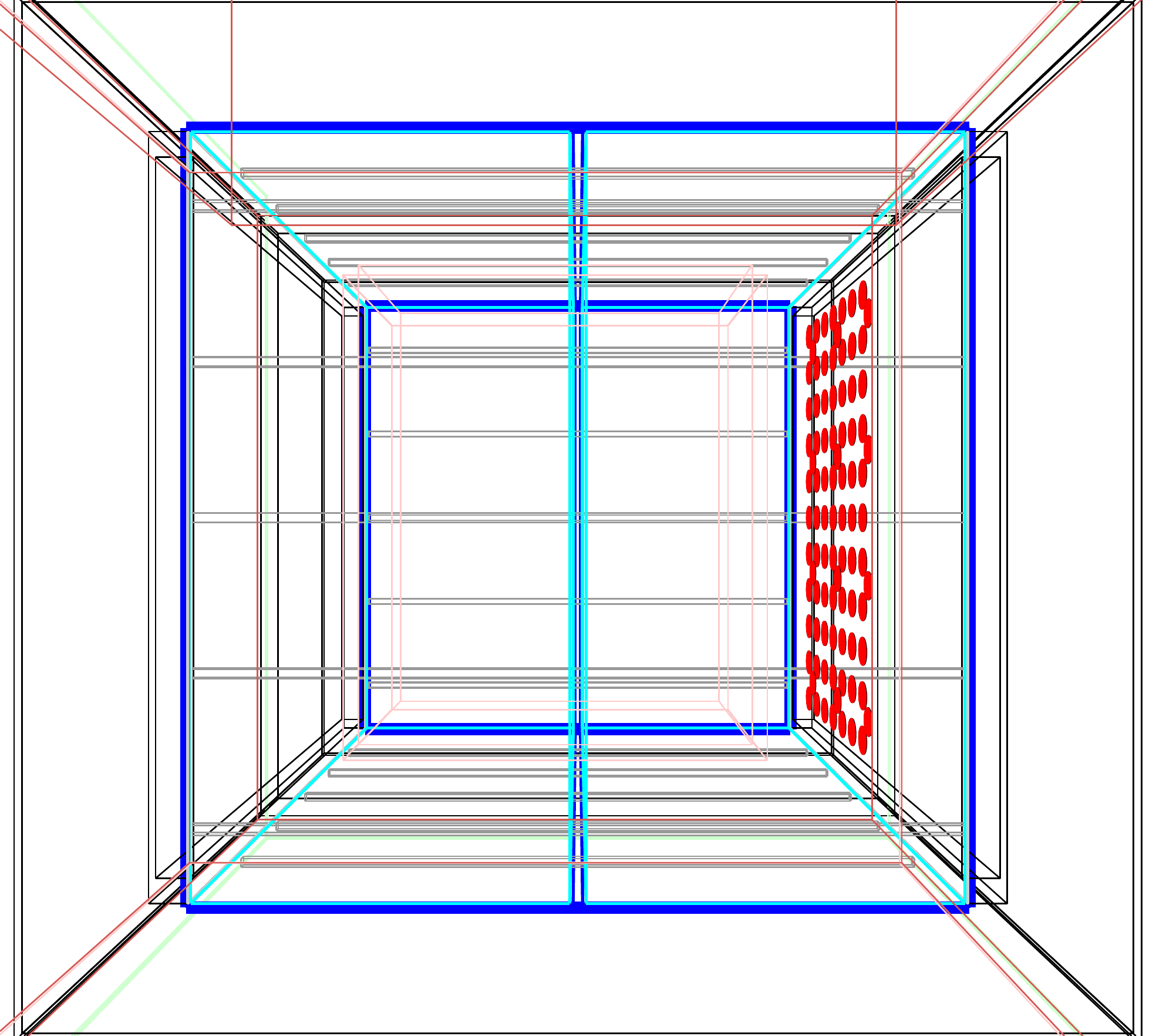}
\caption{Diagram of the SBND geometry used for the generation
  of the optical library. It is visible how the array of PMTs (77
  units) is covering only half of the photocathode plane of the TPC
  used in the simulations. The TPB-coated reflector foils are
  represented in blue, and in cyan the active volume of the TPCs.} 
\label{fig:pmtarray}
\end{figure}

For the generation of the library required by the fast optical
simulation, we divided the active volume of our detector into 3D pixels
or voxels (\emph{voxelization}). To have voxels with 5$\,cm$ in each
dimension (x = 200$\,cm$ drift, y = 400$\,cm$ height and z = 500$\,cm$
length) we defined 40, 80 and 100 voxels in the x, y and z
directions. Subsequently, 400k photons were randomly generated in each of these regions. This was done by a
module in LArSoft that generates a source of an isotropically produced
Geant4 photons from any specified area in the detector. This light was
simulated following a gaussian energy spectrum centered at 9.69$\,eV$
and with a width of 0.25$\,eV$ to imitate the scintillation emission
in liquid argon~\cite{spectrum}. This resulted in an extremely large ($1.28\times10^{11}$) number of photons to be tracked (with
Rayleigh scatterings, reflections, absorptions and wavelength
shiftings) with its consequent consumption of CPU and memory. To
optimize the computational intensity required by this procedure, and
making use of the symmetry in our geometry, we ``switched-off''
(removed from our simulations) the PMTs located in one half of the
photocathode plane (z-y plane). A diagram of the geometry
used for the generation of our library is shown in the
figure~\ref{fig:pmtarray}. From the \emph{reduced} library generated
in this way, by trivial symmetry operations, we could built the
\emph{completed} library with all the needed components.

\subsection{Arrival time distributions}
\begin{figure}[tbp] 
\centering
\begin{minipage}{\textwidth}
 \includegraphics[width=.5\textwidth]{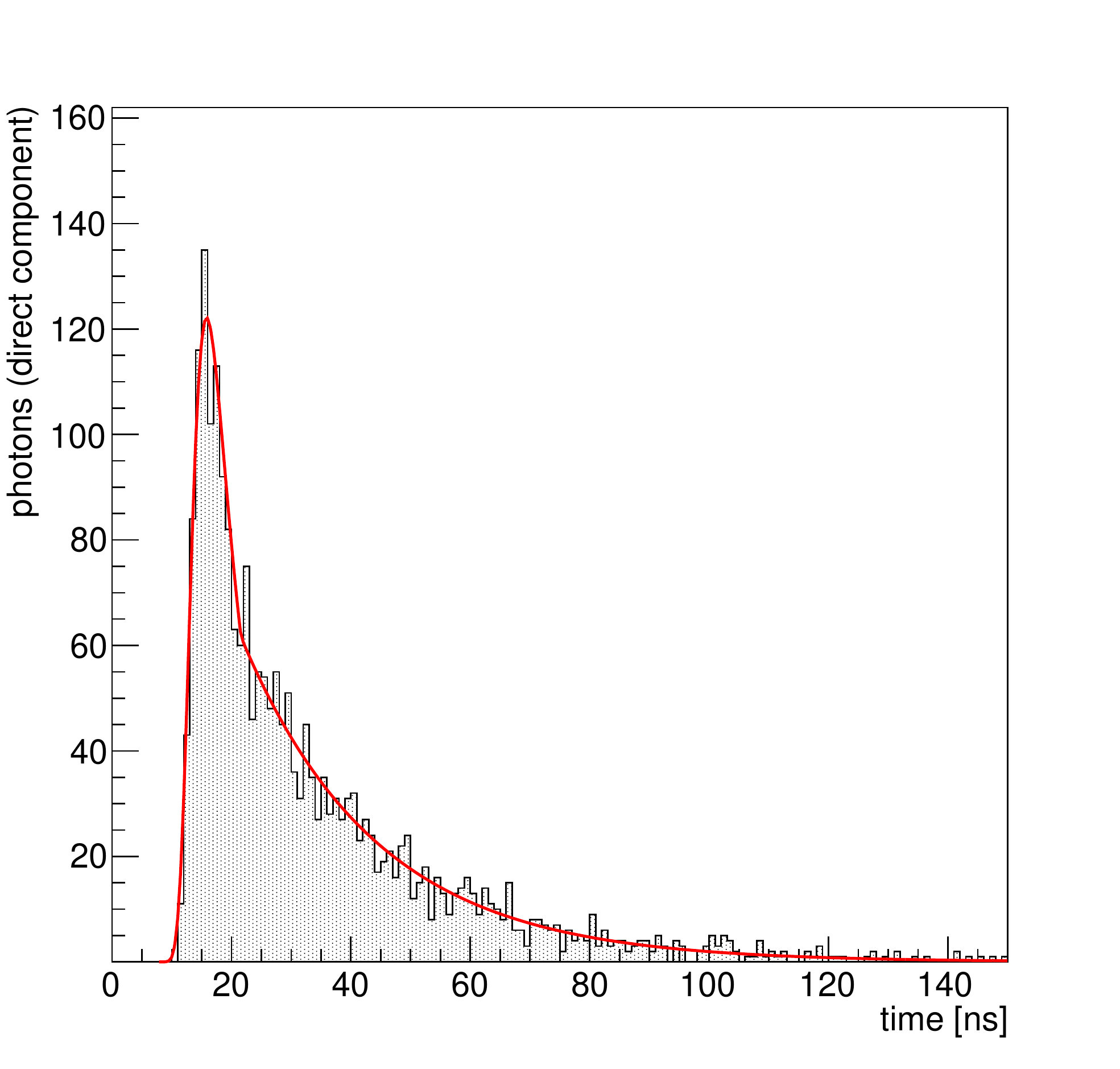}
 \hspace{\fill}
 \includegraphics[width=.5\textwidth]{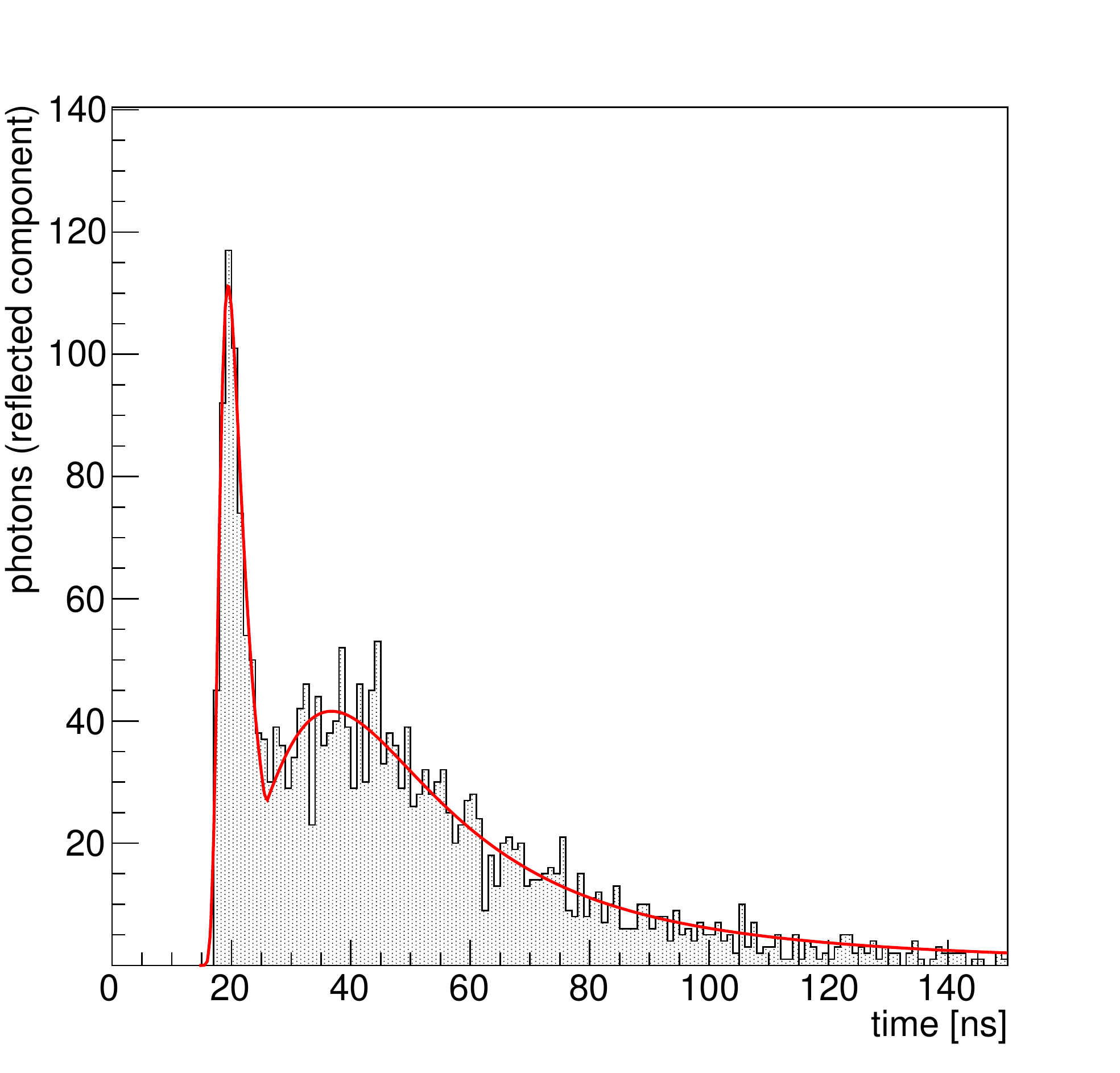}
\end{minipage}
\caption{Two examples of the arrival time distributions recorded by
  one PMT for the direct (left) and the reflected (right) light
  components. The red lines represent the fits of the distributions to
  the functions used to obtain the parametrization.}
\label{fig:fits}
\end{figure}

The scintillation light emission in LAr is governed by its double
exponential decay form (fast and slow components). The optical library
method does not provide the arrival times of the detected photons
which can be affected during direct transport and Rayleigh scattering.
For a "small'' detector like SBND this is a second order effect,
gaining in importance as the path traveled by the photons
until being detected increases. The inclusion of these
distributions into the library would explode the memory. 
Originally, the adopted strategy for the timing was to parametrize
them as a function of the arrival time of the first photon, using only
the information of the PMT location and the scintillation
point. Therefore, no extra information was needed in the optical
library. This works quite well for the case of the direct light.  The
reflected light is more complicated because there is no obvious
(representative) traveled distance for this component. After several
attempts to estimate the minimum average distance (arrival time of the
first photon) for each pair ``scintillation point - PMT'', we decided
to include that piece of information into the library (without
noticeable impact on the memory).


Figure~\ref{fig:fits} (left) shows a typical photon arrival time
distribution for the direct light component. A prompt signal
followed by a diffuse tail are clearly detectable. To model the shape
of these distributions, we have used a gaussian plus an exponential
function (red curve on the figure). An example of the time
distribution for the reflected light component is shown in
figure~\ref{fig:fits} (right). In this case, a sum of two landau
functions has been the model adopted to describe their shapes. The
next step would be to parametrize the parameters of the two selected
functions with the minimum arrival time of a photon in each case\footnote{For the direct light component this is calculated
  simply by the ratio between the voxel-PMT distance and the group
  velocity of the VUV photons generated in the liquid argon
  scintillation.}. At the time of writing this proceeding, these
parametrizations are still work in progress.  

\section{Some preliminary results using the optical
  libraries}\label{sec:results}

\begin{figure}[tbp] 
\centering
\includegraphics[width=1.\textwidth]{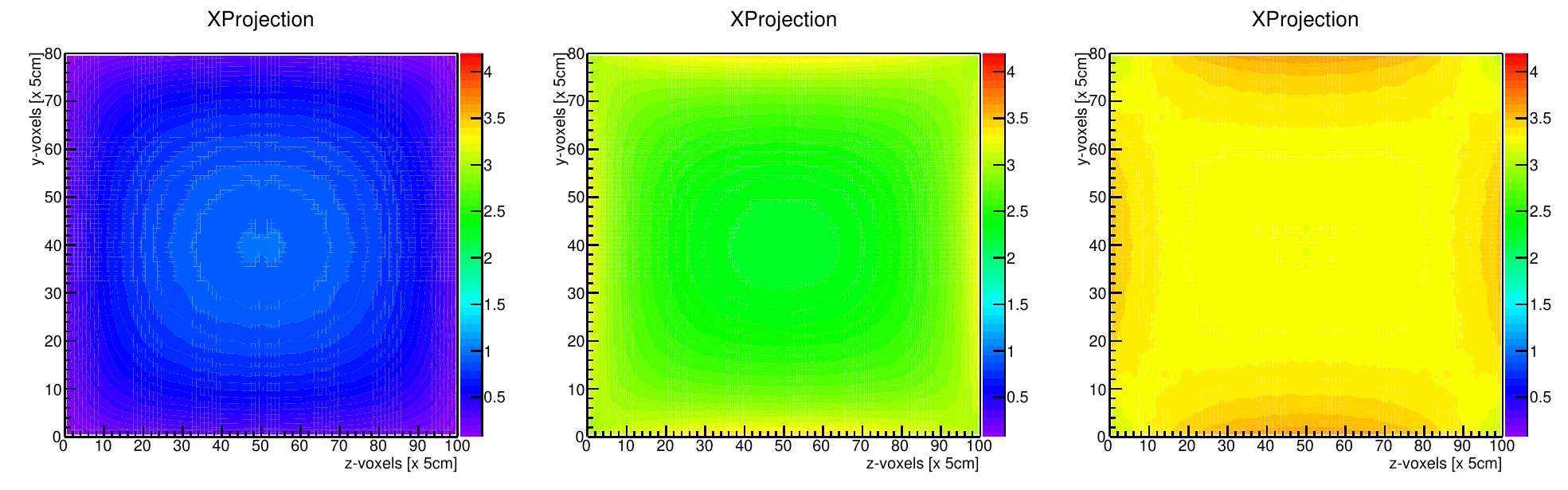}
\includegraphics[width=1.\textwidth]{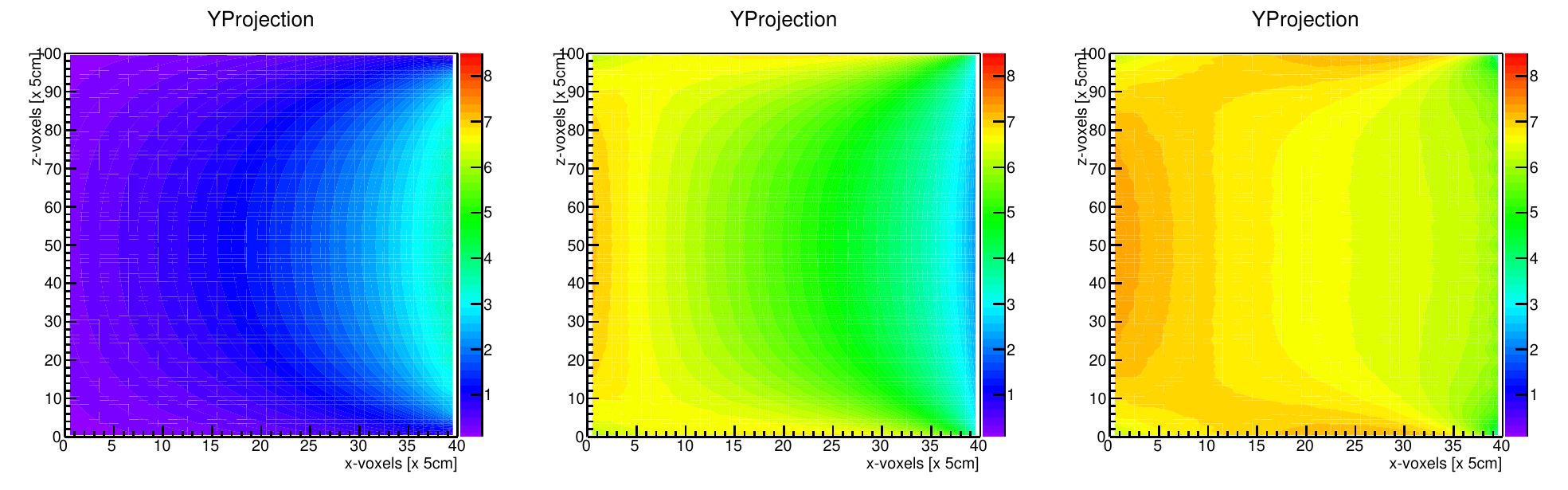}
\caption{Visibility maps projected on the X (drift) and Y (height)
  directions in the TPC active volume: (left) direct light component, (center) reflected light component, (right) total component.}
\label{fig:vismaps}
\end{figure}

One way of testing the system is by using \emph{visibility maps} across the
detector. Figure~\ref{fig:vismaps} shows different examples.
The XProjection and the YProjection 
histograms show the visibility summed for every optical detector
projected down x (drift) and y (height) directions respectively. In other
words, the XProjection represents the total probability of a photon
to reach a PMT summed for every x at each (y, z) position. Thus, they are representations of the photon detection efficiency of
our detector. In figure~\ref{fig:vismaps}, the panels on the right
have been obtained with a LDS consisting of an array of 154 3"
diameter TPB-coated PMTs covering uniformly the photocathode plane
together with TPB-coated reflector foils covering the active volume of
the TPC (2 foils in the xy plane + 2 foils in the xz plane) and the
cathode plane (2 foils in the yz plane). In a system like this, all
recorded signals are a mixing of the direct and the reflected 
components of the generated scintillation light. If we remove the
foils from our system, the only accesible component is the direct
light. This is represented in the panels on the left. In the
middle are shown the results for the reflected component alone. This
latter case would be equivalent to a system with the foils where the PMTs
are not coated by TPB. The main result derived from these figures
is that a PMT-based LDS including reflector foils provides a more efficient
and uniform collection of the scintillation light along the whole
detector volume (by an average factor of the order of 3 and 6,
respectively, relative to our design).

\end{document}